\documentclass[fleqn,usenatbib]{mnras}
\pdfoutput=1

\usepackage{newtxtext,newtxmath}

\usepackage[T1]{fontenc}

\DeclareRobustCommand{\VAN}[3]{#2}
\let\VANthebibliography\thebibliography
\def\thebibliography{\DeclareRobustCommand{\VAN}[3]{##3}\VANthebibliography}

\usepackage{graphicx}	
\usepackage{amsmath}	
\usepackage{braket}

\title[Chemical \& Thermal Structure of NGC5813]{The Chemical and Thermal Structure of the Hot Atmosphere of the Elliptical Galaxy NGC~5813}

\author[D. Chatzigiannakis et al.]{
D. Chatzigiannakis,$^{1}$\thanks{E-mail: d.chatzigiannakis@umail.leidenuniv.nl}
A. Simionescu,$^{2,1,3}$
F. Mernier$^{4,2}$
\\
$^{1}$Leiden Observatory, Leiden University, Niels Bohrweg 2, 2300 RA Leiden, The Netherlands\\
$^{2}$SRON Netherlands Institute for Space Research, Niels Bohrweg 4, 2300 RA Leiden, The Netherlands\\
$^{3}$Kavli Institute for the Physics and Mathematics of the Universe, The University of Tokyo, Kashiwa, Chiba 277-8583, Japan\\
$^{4}$European Space Agency (ESA), European Space Research and Technology Centre (ESTEC), Keplerlaan 1, 2201 AZ Noordwijk, The Netherlands\\
}

\date{Accepted 2022 September 13. Received 2022 September 8; in original form 2022 June 27}

\pubyear{2022}

\begin{document}
\label{firstpage}
\pagerange{\pageref{firstpage}--\pageref{lastpage}}
\maketitle

\begin{abstract}
We present a robust representation of the chemical and thermal structure in the galaxy group NGC~5813 using archival, deep X-ray observations, and employing a multi-temperature spectral model based on up to date atomic line emission databases. The selection of our target is motivated by the fact that NGC~5813 has a very relaxed morphology, making it a promising candidate for the study of the AGN feedback's influence in the intra-group medium (IGrM). Our results showcase a prominent, extended distribution of cool gas along the group's NE-SW direction, correlating with the direction along which the supermassive black hole in the group's central galaxy is known to interact with the IGrM. Our analysis indicates gas being uplifted from the group's centre as the probable origin of the cool gas, although alternative scenarios, such as in-situ cooling can not be explicitly ruled out. Regarding the chemical structure of the IGrM, and unlike previous findings in massive clusters, we find no evidence for recent metal transport by jets/lobes from the central AGN. Instead, elemental abundances remain near Solar on average across the group. The distribution of elements appears to be independent of galactocentric radius, azimuth and the thermodynamics of the gas, suggesting that the IGrM has been efficiently mixed. The large scale uniformity of the abundance distribution implies the presence of complex dynamical processes in NGC~5813, despite its overall relaxed morphology. Past events of extreme AGN feedback or sloshing could be the primary mechanisms behind this.
\end{abstract}




\section{Introduction}

The formation and evolution of the large scale structure of the Universe is directly linked to the production and distribution of heavy elements throughout the cosmos. This is because the same processes that govern galaxy evolution, such as star formation and black hole accretion, also leave a permanent imprint on the observed amount and spatial distribution of chemical elements. As feedback from active galactic nuclei (AGN) is believed to have played a key role in the spread of metals in the early Universe \citep[e.g.][]{Werner13, Urban17, Biffi18b, Mernier18c, Gastaldello21}, it is important to understand in more detail how this transport of chemical elements takes place. Clusters of galaxies, and more specifically their hot atmospheres (the so-called intracluster medium or ICM, \citealt{Mitchell76}), are commonly used to study the interplay between AGN feedback and the chemical evolution of the Universe, as they provide us with the opportunity to examine how metals are distributed over large scales when dissociated from their stellar populations. This has to do with the fact that, due to their large gravitational potential wells, clusters of galaxies effectively retain all elements produced by their member galaxies \citep[e.g.][]{Biffi18b,Mernier18c}.

Regarding the overall chemical composition of the hot atmospheres of galaxy clusters and groups (i.e. the relative abundances of different metals, usually expressed as X/Fe), both the ICM and IGrM appear to be remarkably Solar throughout \citep[e.g.][]{Simionescu15, Mernier17, HitomiCollab18, Mernier18b, Simionescu19}. This is in contrast to the typically non-Solar abundance patterns of the brightest cluster galaxy's (BCG) stellar population. On the other hand, previous studies of the ICM have found strong evidence indicating that the spatial distribution of metals is influenced by mechanisms, such as outflows and jets \citep[e.g][]{Simionescu08,Simionescu09,Kirkpatrick11,Kirkpatrick15}, that are directly linked to AGN feedback. Namely, these studies indicate an increased absolute abundance (with respect to H) along the trajectories of jets and outflows from the central AGN, inferred either via the cluster's morphology, or via the location of known extended radio emission. Those studies have also estimated that the amount of gas being displaced by outflows and jets corresponds to a large fraction of the total hot gas mass surrounding the BCG, making AGN feedback the predominant agent in the spreading of metals from the BCG's stellar population.

Given the prominence of AGN feedback induced metal distribution in clusters of galaxies, it has been suggested that this mechanism could be even more prominent in galaxy groups, considering their shallower gravitational potentials compared to clusters and the fact that this mechanism relies on the uplift of gas. However, studies of the chemical evolution of the IGrM, while limited, remain inconclusive regarding the role of AGN feedback in galaxy groups. Some galaxy groups such as NGC 6051 \citep{OSullivan11} and NGC 4636 \citep{OSullivan05} show clear indications of an increased amount of Fe along the jets and lobes, while others like NGC 4325 \citep{Lagana15} show no evidence of a correlation between enhanced abundance and morphological features. More extreme cases, such as M49 \citep{Gendron17}, show a direct anti-correlation between abundance and AGN feedback features, with a significant under-abundance along the known radio lobes. However such cases might be subjected to effects such as the Fe-bias \citep{Buote99} since the lobes are expected to be multi-temperature in nature \citep[e.g.][]{Simionescu08}. 

In this paper we focus on the bright elliptical galaxy NGC~5813. This is the central dominant member of a subgroup, hereafter referred to as the NGC~5813 group, with an extensive diffuse X-ray emission. While part of the well-isolated NGC 5846 group \citep[$z$=0.006578; e.g.][]{Mahdavi05,Machacek11}, with a projected separation of $\sim 740~\rm{kpc}$, the two show no signs of an interaction between them. NGC~5813 itself has been assumed to be dynamically old \citep[][]{Emsellem07}, with no evidence of a recent major merger in its history, as indicated by the lack of any significant disturbances to its dusty circumnuclear disk \citep[][]{Tran01}. Regarding its atmosphere, the NGC~5813 group has a rather well defined regular morphology, consisting of three collinear pairs of cavities and associated shock fronts, that are products of three distinct outburst events in the central AGN's history \citep[][]{Randall11}. Such cavities often appear as a result of AGN-ICM or IGrM interaction, with the radio lobes, produced by the AGN, displacing the surrounding X-ray gas \citep[e.g.][]{McNamara12,Barai16,Yang19,Gastaldello21}. It has been estimated that an energy of $1.5\times 10^{56}$ and $4\times 10^{57}~\rm{erg}$ has been released by the most and second most recent outbursts, respectively, suggesting that the most recent outburst might still be ongoing. These features are correlated with the group's observed radio emission \citep[][]{Giacintucci11} and can serve as clear indicators of AGN feedback being present. Previous studies of the group have also found moderate turbulent velocities of $\sim 175~\rm{km s^{-1}}$ and a 3D Mach number of the order of 0.4, using resonant scattering \citep[e.g.][]{dePlaa12,Ogorzalek17}, further indicating the presence of such a mechanism. These characteristics make NGC~5813 a very promising candidate for the study of the effects AGN feedback can have on the distribution of elements in a low-mass system. Additionally, this target has some of the deepest Chandra data of any galaxy group and, similar to M49, has been found to have an explicit anti-correlation between its metal abundance and the location of its radio lobes \citep{Randall15}. That work already suggests that this result is dependent on the Fe-bias; employing a two-temperature fit brings the abundance in the region of the extended radio lobes in better agreement to their surrounding medium, although the central region appears to remain under-enriched.

In this work, we aim to reanalyse the archival Chandra data for NGC~5813, with two important differences compared to previous studies. First, we will use a multi-temperature spectral model throughout, in order to map the IGrM's metal and multi-temperature distribution in greater spatial detail than previous work. Second, we will use the most up to date atomic line emission models, which have evolved significantly over the last years, in particular when it comes to describing the Fe-L complex that dominates the emission at the low temperatures corresponding to the IGrM \citep[][]{Gu20,Gastaldello21}. In Section \ref{sec:data} we present our initial data analysis, focusing on data preparation and spectral model initialisation. The results of our spatially resolved spectroscopy are presented in Section \ref{sec:analysis}. We discuss our findings in Section \ref{sec:discussion}, while a summary is presented in Section \ref{sec:conclusions}. For the purposes of this work, we assume a standard $\Lambda$CDM cosmology ($\Omega_m=0.3$, $\Omega_\Lambda=0.7$ and $H_0=70~\rm{km~s^{-1}Mpc^{-1}}$) which, for the redshift of NGC~5813, corresponds to a scale of $0.15~\rm{kpc}/"$. Uncertainties are given at 68\% confidence intervals (1$\sigma$), unless otherwise stated. Abundances are estimated using the \citet{Asplund09} Solar abundance reference tables.

\section{Data Analysis} \label{sec:data}
\subsection{Observations and Data Reduction}
The \textit{Chandra} observations available for NGC~5813 are presented in Table \ref{tab:observations}. All data were reprocessed from the level 2 event files using Chandra Interactive Analysis of Observations (CIAO) software and CALDB 4.9.6. Background flares have been removed by running CIAO's \textit{deflare} routine on the extracted lightcurves, applying a high energy filter (7-12~keV) on each event file, and employing the \textit{lc\_clean} algorithm. We note that no periods of strong flares were found, consistent with previous studies using the same set of observations \citep{Randall15}. Blank sky images were also generated for each observation, using CIAO's \textit{blanksky} function. The background images have been normalized to match the 9-12~keV particle count rate of each respective observation. As the observations have been performed in VFAINT mode, all events flagged as a "potential background event" have been filtered out.

\begin{table}
    \centering
    \begin{tabular}{l l c}
    \hline \hline
    Obs ID & Date Obs & Cleaned Exposure (ks)\\
    \hline
    5907 & 2005 Apr 2 & 49.00\\
    9517 & 2008 Jun 5  & 100.06\\
    12951 & 2011 Mar 28 & 74.95\\
    12952 & 2011 Apr 5 & 144.91\\
    12953 & 2011 Apr 7 & 32.16\\
    13246 & 2011 Mar 30  & 45.60\\
    13247 & 2011 Mar 31 & 36.22\\
    13253 & 2011 Apr 8  & 119.54\\
    13255 & 2011 Apr 10  & 43.96\\
    \hline \hline
    \end{tabular}
    \caption{\textit{Chandra} X-Ray Observations}
    \label{tab:observations}
\end{table}

\subsection{Image Analysis}
\begin{figure*}
    \centering
    \includegraphics[width=0.5\linewidth]{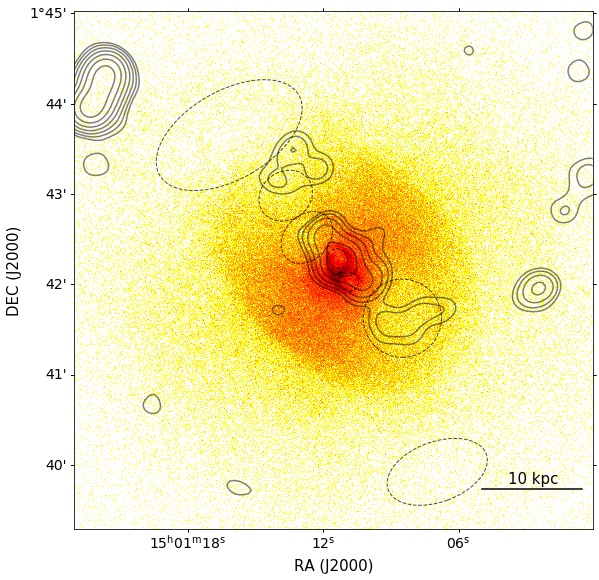}
    \caption{Exposure-corrected, background subtracted, 0.3-3~keV \textit{Chandra} image, smoothed with a $\sigma=1.5"$ Gaussian kernel. Point sources were removed and interpolated using a Poisson method. Contours indicate \textit{GMRT} 236 MHz emission (FWHM=16.0"$\times$ 13.0", p.a. $0\deg$; $1\sigma$=$300\mu\text{Jy}~ \text{beam}^{-1}$), spaced by a factor of 2 starting from $0.9m\text{Jy}~ \text{beam}^{-1}$ (radio data adopted from \citet{Giacintucci11}). Dashed contours indicate the location of several generations of cavities, as described in Table 4 of \citet{Randall15}}
    \label{fig:smoothed}
\end{figure*}

We merged all available observations to the same tangent point and corrected for the exposure and background maps corresponding to each data set. Bright sources have been identified using \textit{wavdetect} and then, in order to both avoid sampling background X-ray sources during our spatially resolved spectral analysis and highlight NGC~5813's structure, they have been adequately masked. All masked regions have been filled in using a Poisson distribution with a mean equal to the counts of a local-background annulus, using CIAO's \textit{dmfilth} function. The smoothed, point-source removed image is shown in \autoref{fig:smoothed}.\\ 

Based on these images, we recover a structure similar to that of previous NGC~5813 studies \citep[e.g.][]{Randall15,Randall11}, consisting of three distinct generations of cavities, appearing in pairs along the SW-NE axis. For each cavity pair, an extended elliptical edge surrounding it is evident, indicating the shock fronts produced during each generation's expansion phase. The recovered structure appears to be regular, consistent with the notion of a history of "outbursts" from the central AGN, responsible for the creation of the cavities and their associated shock fronts, as previously postulated.

\subsection{Spectral Analysis}
For the purposes of our analysis, we first define spectral extraction regions using the contour binning technique as described in \citet{Sanders06}. This algorithm generates spatial bins following morphological features, such as rims or cavities, according to their surface brightness contrast. For our mapping we chose a minimum signal-to-noise ratio of 50, as the best compromise between spectral and spatial resolution. A geometric constraint of 1 has also been applied to the binning process, restricting the elongation of our regions, in favor of obtaining a more localized view of the IGrM's physical conditions.

Our spectra and their respective backgrounds have been extracted using CIAO's pipeline with the \textit{specextract} function. The background spectra have similarly been scaled to match the observed count rate in the 9-12~keV band. This correction factor was calculated for each ObsID using a box region, spanning the entirety of the field of view (FOV). All subsequent background spectra, such as the ones extracted for individual spatial regions, have then been scaled using the corresponding correction factors.

All of our spectra were fitted in the $0.6-3.0 ~\rm{keV}$ band. The energy range was chosen to include the Fe L-complex, which is the main descriptor of the properties of galaxy groups, while the lower limit was explicitly chosen to avoid calibration uncertainties around the O edge. For our analysis we will be using XSPEC's AtomDB 3.0.9 atomic database, unless otherwise specified, primarily applying two models: (i) a single temperature model, consisting of an absorption component (PHABS) applied on a single thermal component (APEC), and (ii) a double temperature component, consisting of two distinct APEC components with a common PHABS applied to both of them. Regarding the absorption, we have chosen a hydrogen column density fixed at the Galactic value of $N_H=4.37 \times 10^{20} ~ \rm{cm}^{-2}$  \citep[]{Kalberla05}. As for the target's redshift, we adopt a fixed value of $z=0.0066$ for the APEC components. For the double temperature model, additional conditions have been put in place, namely coupling the abundance parameters of its two APEC components in order to constrain the fit and initializing the temperature of the second thermal component (cool gas) at about half of the first thermal component's (hot gas) temperature.

\begin{figure*}
    \centering 
    \includegraphics[width=0.45\linewidth,height=6.5cm]{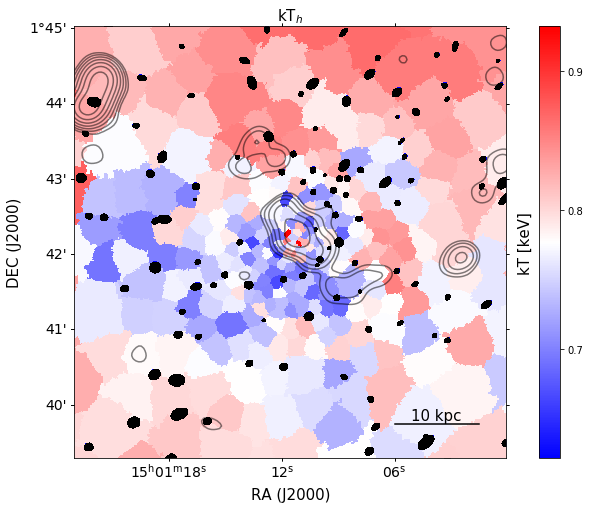}
    \includegraphics[width=0.45\linewidth,height=6.5cm]{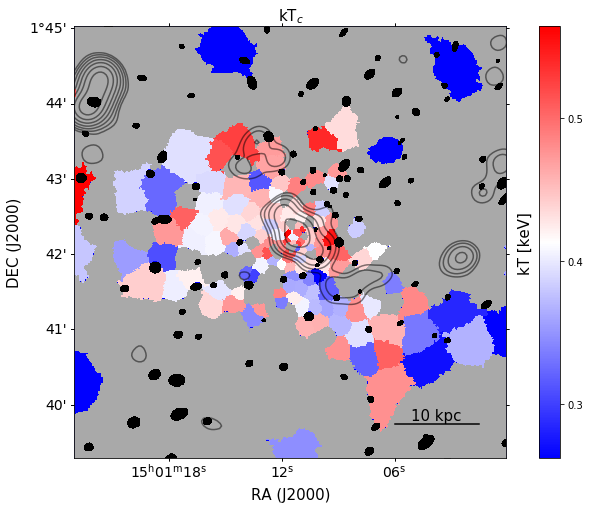}
    \medskip
    \includegraphics[width=0.45\linewidth,height=6.5cm]{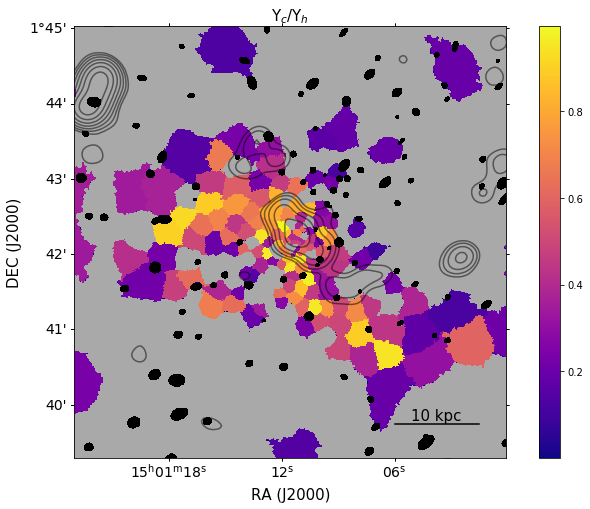}
    \includegraphics[width=0.45\linewidth,height=6.5cm]{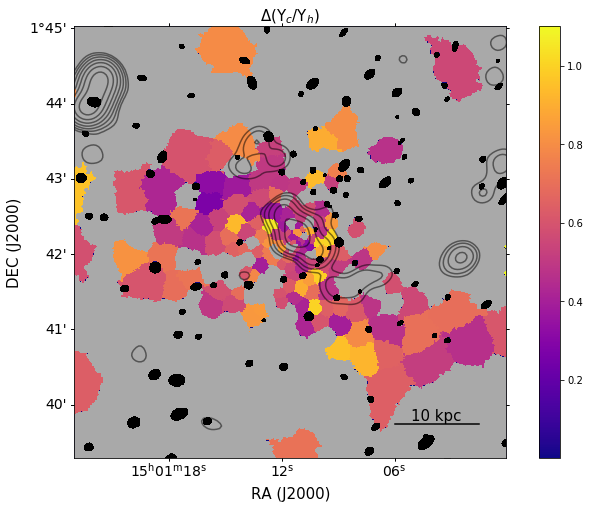}
\caption{Contour binned maps for the hot gas temperature (top left), cool gas temperature (top right), the normalisation ratio of the two thermal components (bottom left) and its respective relative error (bottom right). Grey regions indicate the absence of a second thermal component and black contours indicate \textit{GMRT} 236 MHz emission, similar to \autoref{fig:smoothed}}
\label{fig:temp}
\end{figure*}

Our fitting strategy proceeds as follows: When fitting a single temperature model, all relevant free parameters (i.e. temperature, elemental abundances and normalisation) are fitted at the same time. However, for the double temperature model, due to the larger number of free parameters, it is more difficult for the fitting algorithm to minimize all parameters simultaneously. Therefore, we choose to minimize all free parameters gradually. Initially, we only allow the relevant parameters of the first temperature component to vary, while freezing the second component's normalization ($\text{Y}_2$), setting it to a lower value compared to the first component's normalisation ($\text{Y}_1$) ($\rm{log}\frac{\text{Y}_2}{\text{Y}_1}\simeq -4$), and temperature ($kT_2$), which is set to $0.35~\rm{keV}$, approximating a single temperature fit. Subsequently, we thaw $\text{Y}_2$ allowing it to minimise under our temperature distribution assumption. Finally we allow the $kT_2$ to vary as well. If at any point, but more specifically during the last two fitting steps, a component's normalization or temperature was unconstrained, the region is flagged as a single temperature one.

\section{Spatially Resolved Spectroscopy} \label{sec:analysis}
\subsection{Spectroscopic Maps}

In this section we present the results of our double temperature spectral fitting, unless the spectrum has been flagged as a single temperature one, in which case we use the single temperature fit results instead. As expected, for the cases where the second temperature component was not significant, the effects of the Fe-bias were also negligible, with both models agreeing on the elemental abundance estimates. In the following sections we will refer to the thermal component with the highest temperature as the hot component, with the other component as the cool one. We note that for all cases the hot thermal component is the more significant contributor, as indicated by its higher normalisation value compared to the respective cool component. For the cases where only one thermal component was found, the single thermal component is regarded as the hot one, assuming a cool component with a normalisation of 0.

Regarding the hot temperature component, seen in \autoref{fig:temp}, the IGrM shows only small variations around a mean temperature of $kT_h \simeq 0.78 \pm 0.08~\rm{keV}$, comparable to the temperature of other groups of galaxies \citep[e.g.][]{Rasmussen07}, with the presence of a small radial gradient. The temperature of the cool gas also does not appear to vary significantly, with an average of $kT_c \simeq 0.41\pm 0.08~\rm{keV}$ based on our bins, however its spatial distribution exhibits a clear asymmetry. The bins along the SW-NE axis, associated with the location of the lobes, systematically exhibit a second temperature component.

\begin{figure*}
    \centering 
    \includegraphics[width=0.45\linewidth,height=6.5cm]{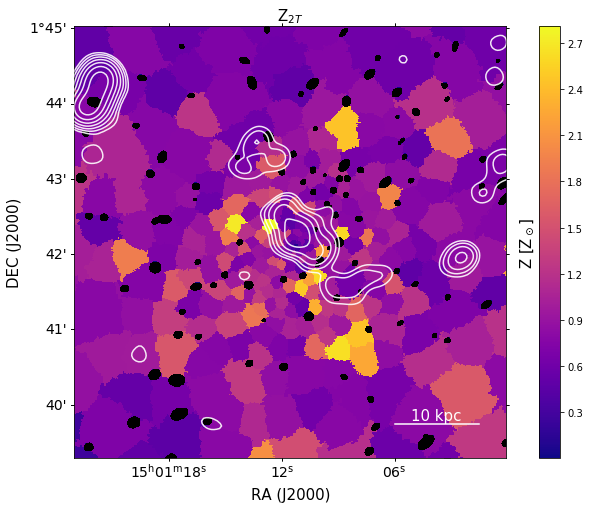}
    \includegraphics[width=0.45\linewidth,height=6.5cm]{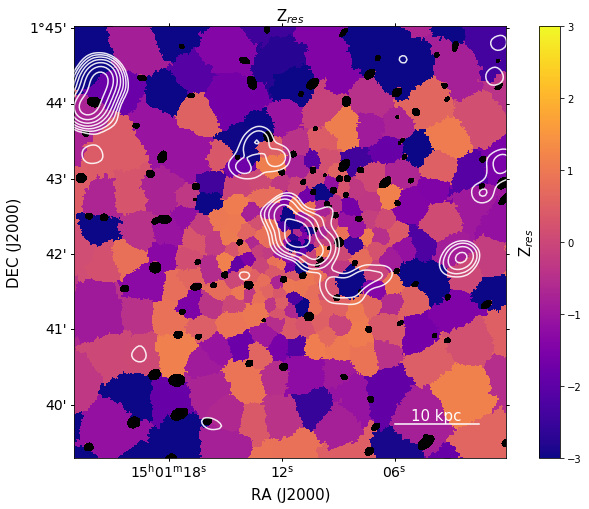}
\caption{Contour binned maps for the overall abundance (in Solar units; left panel) and its residuals compared to a flat abundance profile set at $0.87~\text{Z}_\odot$ (right). White contours indicate \textit{GMRT} 236 MHz emission, similar to \autoref{fig:smoothed}}
\label{fig:abund}
\end{figure*}

The contribution of the second thermal component in the spectra of the bins along the SW-NE direction appears to be prominent, as indicated by the normalisation ratio of the two components, seen in \autoref{fig:temp}. The distribution indicates a large amount of cool gas extending all the way to the intermediate cavity in the north and almost all the way to the outer cavity in the south. Along the two off-lobe directions on the other hand, while a prominent contribution from the cool gas is found in the group's centre, the ratio has a far steeper radial decrease, approaching 0 (i.e. single temperature) within $\sim 10 ~\rm{kpc}$. Overall, the majority of the multi-phase structure appears to be concentrated around the inner and intermediate cavities. However, in the north, the multi-temperature structure appears to be turning eastwards, leading to the extensive SE feature, deviating from the group's cavity alignment.

Despite the strong azimuthal asymmetry in the cool gas distribution, the Fe abundances, shown in \autoref{fig:abund} show no evidence of an abundance enhancement across any direction, with the distribution remaining almost uniform with both radius and azimuth. Our results indicate an average abundance, weighted by the signal-to-noise ratio of each bin, of $Z=0.87\pm0.36~Z_\odot$.The notable scatter of 0.36, computed as the standard deviation of all values in the map, is driven primarily by a number of low signal to noise measurements which tend to have unusually large values. However, the overall distribution remains consistent with a flat abundance profile when examining the spatial abundance residuals with respect to a uniform abundance distribution set at the average value of $\text{Z}_\text{exp}(r)=0.87~\text{Z}_\odot$. Said residuals are calculated for every bin as:
\begin{equation}
    \text{Z}_\text{{res}}=\frac{\text{Z}-\text{Z}_\text{exp}(r)}{\Delta\text{Z}}
\end{equation}
where Z is the measured abundance and  $\Delta\text{Z}$ the statistical error on the measurement, based on our model's fit. The residuals deviate from the average by less than 1$\sigma$ for $\sim57\%$ of our bins, while $79\%$ deviate less than 2, indicating the lack of an abundance enhancement across the galaxy group.

\subsection{Azimuthally Resolved Analysis}\label{azimuthal_analysis}
Despite the presence of an extended cool gas distribution, the elemental abundance, as indicated in the previous section, appears to show no evidence of an enhancement. One possible explanation of this could be the large scatter, and associated uncertainties, of the abundance estimates of the individual bins. Another possibility could be the fact that what we assume to be cool gas might not be thermal in nature, but we will choose to examine this in \ref{hybrid}. In order to address the first point, we choose to rebin our observations along annular sectors defined by the observed cool gas distribution. While worsening our spatial resolution, this approach reduces the uncertainties of our fitting output parameters. At the same time, the azimuthal binning still allows us to directly examine the effect of the cool gas by comparing the two lobe directions (i.e. SW and NE), where a prominent cool gas contribution is found, with the two off-lobe directions (SE and NW), including regions predominately found to lack such a component.

\begin{figure}
\includegraphics[width=\linewidth]{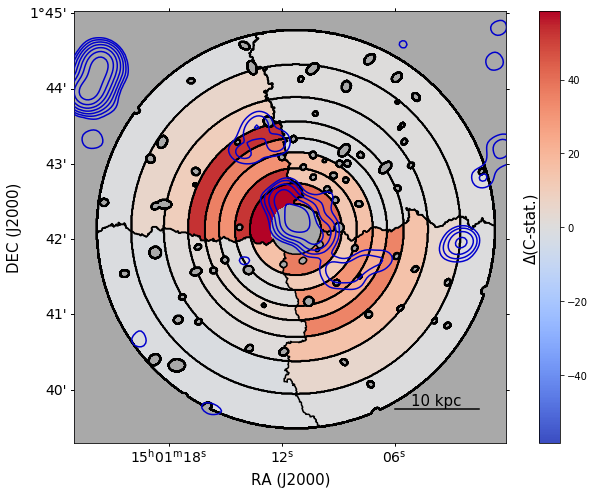}
\caption{Map of the fit's C-statistic difference for each bin, using the hybrid and double temperature models ($\Delta(\text{C-stat.})=\text{C-stat}_{\text{hybrid}}-\text{C-stat}_{\text{double~temp.}}$). Blue regions indicate a smaller C-statistic value, i.e. a better fit for the hybrid model, and vice versa for the red ones. Blue contours indicate \textit{GMRT} 236 MHz emission, similar to \autoref{fig:smoothed} }
\label{fig:pow_cstat}
\end{figure}

\begin{figure*}
    \centering 
    \includegraphics[width=0.45\linewidth,height=5.5cm]{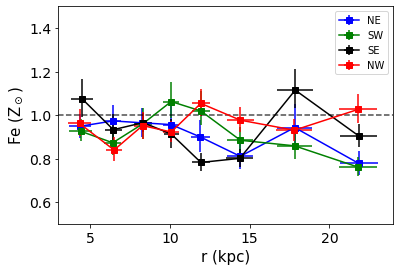}
    \includegraphics[width=0.45\linewidth,height=5.5cm]{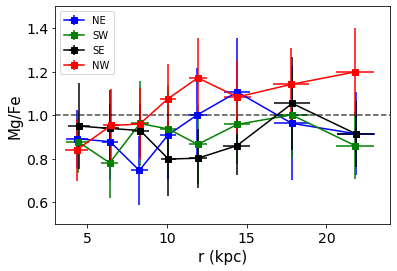}
    \medskip
    \includegraphics[width=0.45\linewidth,height=5.5cm]{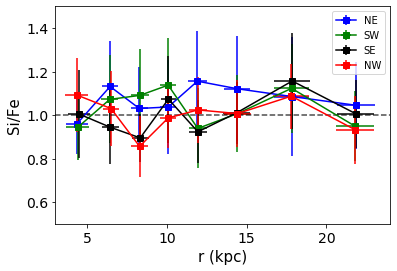}
    \includegraphics[width=0.45\linewidth,height=5.5cm]{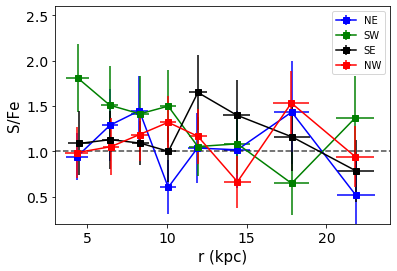}
\caption{Radial profiles for the elemental abundances of Fe (top left), Mg (top right), Si (bottom left) and S (bottom right) for all 4 directions based on our two-temperature model. The dashed line indicates Solar abundance, for the case of Fe, and a ratio of 1 with respect to Fe for the rest of the elements.}
\label{fig:abund_radial}
\end{figure*}

Regarding the azimuthal binning, the four directions are chosen to follow the edges of the observed cool gas distribution from \autoref{fig:temp}. Regarding the radial binning, each annulus has been chosen to have a thickness of at least $1.5"$, in order to avoid photon leaking from nearby regions, while demanding a minimum number of 20,000 counts in the $0.6-3~\rm{keV}$ range, as a means to ensure high quality fits. In the interest of directly comparing the radial properties of the IGrM, we force the annuli sectors in all four azimuthal directions to have the same thickness at any given radius, which leads to some azimuths having an excess of counts above the minimum, especially the central ones. An example of our binning can be seen in \autoref{fig:pow_cstat}. We also note that, with this binning, we choose to ignore the central region ($r<3~\rm{kpc}$) containing the inner cavities, central AGN and shock fronts. We examine this region separately in \ref{central}, opting for a feature-driven binning instead.

\subsubsection{Nature of the Multi-Temperature Structure}\label{hybrid}
One possible explanation behind the lack of an abundance enhancement could be that the second component, which we have so far assumed to be indicative of cool gas, might in fact not be thermal in nature. As previously demonstrated, the observed directional asymmetry in the distribution of the cool gas correlates strongly with the direction of the group's detected radio emission \citep[e.g.][]{Giacintucci11,Birzan20}, the latter implying the presence of relativistic non-thermal electrons allowing for the possibility that the second component is a product of non-thermal processes, such as inverse Compton scattering, instead.

In order to test this hypothesis,  we define a new model in XSPEC, hereafter referred to as a hybrid thermal model. This consists of a thermal component and a power-law component, instead of our previously used double temperature model, in order to simulate the non-thermal contribution, applying an absorption component on both.  If the emission is non-thermal, the hybrid thermal model should outperform the double temperature one, by systematically having a smaller goodness of fit (G.O.F.) out of the two. As both models have the same number of 1462 degrees of freedom (D.O.F.), we can directly compare the C-statistics of the two models. As \autoref{fig:pow_cstat} indicates, the hybrid thermal model across the two lobe directions is systematically worse than the double temperature model confirming that the emission detected is indeed thermal in nature. The fact that the additional power-law component does not fit the data as well as a double temperature model can be easily understood if the multi-temperature constraints are driven by the shape of the Fe-L bump, rather than the shape of the continuum.

\subsubsection{Radial Profiles ($r>3~\rm{kpc}$)}
Since, as we have demonstrated, an underlying cool gas distribution is present across the IGrM, we reexamine its elemental abundance distribution to search for weaker evidence of an abundance enhancement that may have been missed in the spatial maps. In \autoref{fig:abund_radial} we present the radial profiles of Fe, and the abundance ratios of Mg, Si and S with respect to Fe, across our four previously defined sectors. We note that, due to our new binning scheme, we are able to recover the abundances of Fe, Mg and Si, with a significance of $3\sigma$, while S is detected at $\sim 2\sigma$ in the outer bins. As found earlier, the Fe abundance remains similar in all 4 directions across the galaxy group, despite their different physical properties. More interestingly, the profiles lack a radial gradient, further indicating a spatial homogeneity across the entire group. This behaviour appears to be contradicting what previous studies of relaxed clusters/groups \citep[e.g.][]{dePlaa17,Simionescu09, Sun12, Mernier17, Lovisari19, Gastaldello21} have found, with the Fe abundance exhibiting a prominent central enhancement peak that drops towards higher radii. Regarding the rest of the elemental abundances, Mg seems to indicate an increase with radius in the NW direction but, due to the uncertainties involved, the trend can be approximated with a constant ratio, remaining within error bars in agreement with the other three directions. Both Si and S show no particular trends with radius in either direction, suggesting an even spread of said elements across NGC~5813.

\subsubsection{Central Region ($r<3~\rm{kpc}$)}\label{central}
\begin{figure}
    \centering 
    \includegraphics[width=\linewidth]{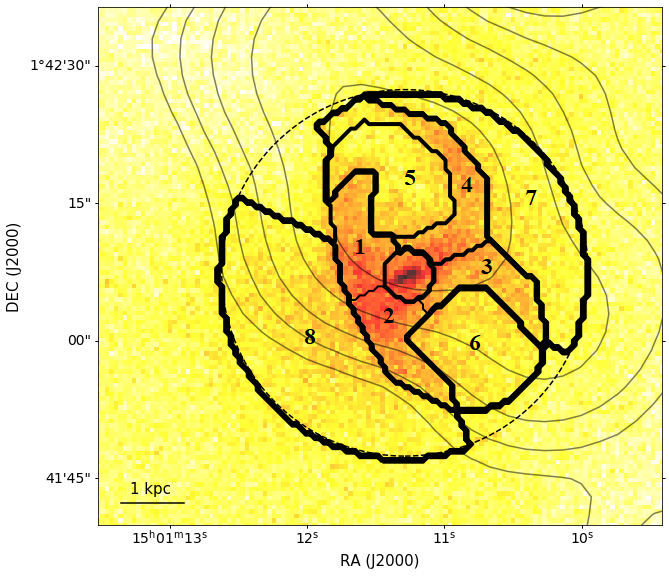}
\caption{Schematic of the central $3~\rm{kpc}$ binning. Numbers indicate the bin index of a given region. The dashed line shows the edges of the central region ($r=3~\rm{kpc}$) while background contours indicate \textit{GMRT} 236 MHz emission, similar to \autoref{fig:smoothed}}
\label{fig:central}
\end{figure}

In the central region ($r<3~\rm{kpc}$), the cool gas distribution extends across all azimuths. Therefore, instead of a directional analysis of the elemental abundance, we instead choose to examine regions drawn along morphological features, namely the 4 rims of the inner cavities (bins 1-4), the two inner cavities themselves (bins 5 and 6) and finally the surrounding atmosphere in the NW and SE directions (bins 7 and 8). The region binning for the central $3~\text{kpc}$ can be found in \autoref{fig:central}.

We note that certain pixels inside the central region have not been accounted for. In particular, both towards the SW and NE, we are likely seeing `channels' connecting to the intermediate generation of cavities farther out in radius. The properties of these regions may be expected to differ from those of the chosen bins, although the count number is insufficient to study them independently. The central AGN has also been masked, as to avoid any contamination in our spectra.

\begin{figure*}[b]
    \centering 
    \includegraphics[width=0.45\linewidth,height=5.0cm]{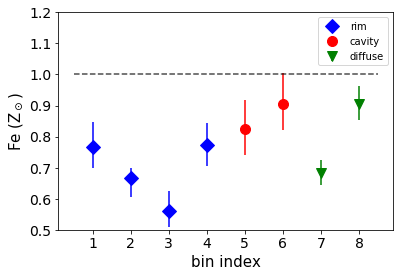}
    \includegraphics[width=0.45\linewidth,height=5.0cm]{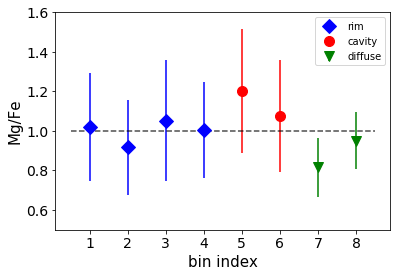}
    \medskip
    \includegraphics[width=0.45\linewidth,height=5.0cm]{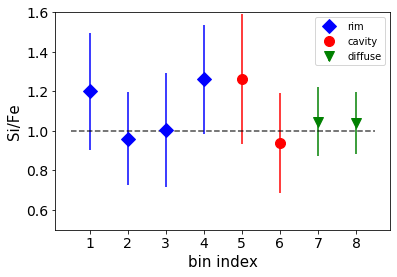}
    \includegraphics[width=0.45\linewidth,height=5.0cm]{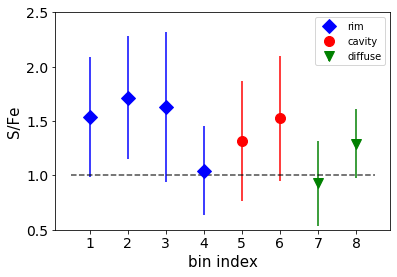}
\caption{Elemental abundances of Fe (top left), Mg (top right), Si (bottom left) and S (bottom right) for each central bin based on our two-temperature model. The dashed line indicates Solar abundance, for the case of Fe, and a ratio of 1 with respect to Fe for the rest of the lines.}
\label{fig:abund_central}
\end{figure*}

In \autoref{fig:abund_central} we present the elemental abundances for the central morphological features of NGC~5813. Overall, the central region appears to be below Solar, with an average Fe abundance of $0.76\pm 0.11~Z_{\odot}$, further highlighting the lack of a central enhancement peak. When it comes to the two surrounding atmosphere regions, an abundance asymmetry is apparent, with the SE direction (bin 8) appearing metal-richer than its NW counterpart. Perhaps more interestingly, this asymmetry, unlike the larger scale one found along the SE-NW axis, appears to not be related to the cavity alignment. For the other elements (Mg, Si, S), the relative abundance with respect to Fe remains in good agreement with the Solar composition.

Overall, the derived abundances further highlight the absence of any abundance asymmetries across the galaxy group, indicating metals being very well mixed in the IGrM. The only deviation from a such uniformity in the elemental distribution has been found in the centre of the group, suggesting a possible scale dependence of the elemental mixing, with the mixing taking place primarily at larger scales. We also point out that our results remain invariant under different parameterisations of our models and different atomic libraries, as can be seen in greater detail in Appendix \ref{systematics}.

\section{Discussion}\label{sec:discussion}
\subsection{Origin of the Multi-Temperature Structure}
As previously shown, the cool gas distribution indicates a strong directional asymmetry, appearing almost exclusively in the center of the galaxy group and along its radio lobes, where the cavities are present. We have confirmed that this component is thermal in nature, based on the spectral shape (seen in \autoref{fig:pow_cstat}). A natural follow-up is to examine the possible origins of the cooler plasma phase. In the following paragraphs we will discuss the two most likely scenarios that can explain the presence of this cool gas and compare them with its observed distribution. These scenarios are (i) clumps of gas cooling in place (in situ cooling) due to thermal instabilities in the IGrM, and (ii) cooling due to gas being uplifted from the group's centre during 'outburst' events in NGC~5813's history.

First we consider the in-situ cooling scenario. Previous studies have indicated that the hot gas in the intra-cluster medium (ICM) is capable of cooling in place via classical thermal instabilities \citep[e.g.][]{Hu85,Heckman89,Cavagnolo08,Rafferty08,Voit08}. This scenario is appealing if one considers that the generations of cavities found in NGC~5813 can be an effective way to trigger such instabilities in the IGrM \citep[e.g.][]{Churazov01,McNamara16}. Either due to the central AGN's outburst itself, or as a result of the cavities being pulled outwards, fluctuations are introduced in the hot gas that, if able to grow, can effectively cool down. Such an interpretation can explain why cool gas is predominately found in the two lobes, around the cavities, rather than in the less disturbed surrounding atmosphere directions, effectively addressing the observed directional asymmetry.

\begin{figure*}
    \centering 
    \includegraphics[width=0.45\linewidth]{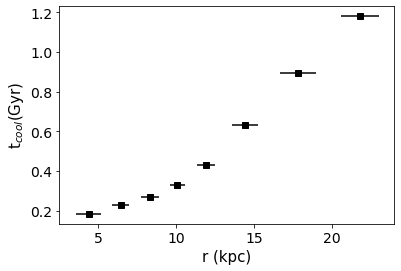}
    \includegraphics[width=0.45\linewidth]{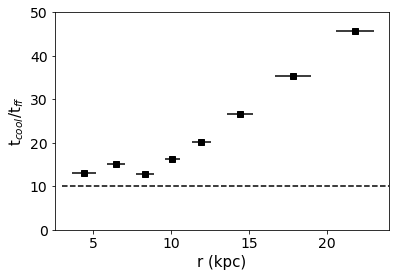}
\caption{Azimuthally averaged radial profiles of cooling time and cooling to free-fall time ratio. The dashed line indicates the limit, below which, thermal instabilities are expected to grow non-linearly.}
\label{fig:t_cool}
\end{figure*}

One way to test whether thermal instabilities can grow large enough to effectively induce an in-situ cooling of the hot gas, is to compare the cooling time of the gas ($t_{\rm{cool}}$) to its free-fall time ($t_{\rm{ff}}$). 
Numerical simulations \citep[e.g][]{McCourt12,Sharma12} have indicated that, for cases where $t_{\text{cool}}/t_{\text{ff}}$ falls below $\sim 10$, any thermal instabilities introduced to the hot plasma are able to grow non-linearly. While this is not an absolute limit, it can serve as a reliable indicator of in-situ cooling, considering the large uncertainties involved in the derivation of the ratio. For our analysis we calculate the cooling time of the hot gas component for each radial bin as:
\begin{equation}
    t_\text{cool}=\frac{\frac{3}{2}(n_e+n_i)k_BT_h}{n_en_i\Lambda(T)}=\frac{3k_BT_h}{n_e\Lambda(T)}~~~~,
\end{equation}
where $n_e$ and $n_i$ are the electron and ion number densities, assuming that $n_e\sim n_i$, and $\Lambda(T)$ is the cooling function of \citet{Schure09} for each bin's abundance. The free-fall time is simply derived as:
\begin{equation}
    t_\text{ff}=\sqrt{\frac{2r}{g}}
\end{equation}
with the gravitational acceleration $g$ derived from the group's pressure profile as:
\begin{equation}
    g=-\frac{1}{\rho}\frac{dp}{dr}=-\frac{1}{n_em_H\mu}\frac{dp}{dr}
\end{equation}
assuming a mean atomic weight $\mu=0.62$. For this calculation, we used the deprojected pressure profiles for NGC~5813 already provided by \citet{Randall15}. In \autoref{fig:t_cool} we present the azimuthally averaged $t_{\text{cool}}$ and $t_{\text{cool}}/t_{\text{ff}}$ radial profiles. All values of the ratio appear to be above the $\sim 10$ threshold. In the central 10 kpc, the derived ratio is just slightly higher than 10; given uncertainties related to azimuthal variations and the projected geometry of the source, we cannot exclude that in reality $t_{\text{cool}}/t_{\text{ff}}$ may drop below the threshold where thermal instabilities become likely. However, at larger radii, $t_\text{{cool}}/t_{\text{ff}}$ becomes much larger than 10. Therefore, the large amounts of cool gas found leading up to the outer cavities ($r\sim 20~\rm{kpc}$) cannot be explained by the in situ cooling scenario.

The second scenario we examine is the possibility that the cool gas distribution is a product of gas being uplifted from the group's centre. Previous studies of clusters of galaxies \citep[e.g.][]{Nulsen05,Wise07,Simionescu08,Simionescu09} have indicated that uplifting gas from a central high-pressure region to larger radii, where the pressure is significantly lower, can be an effective way to adiabatically cool it. Considering that our observed cool gas distribution correlates strongly with the direction of possible outflows, as indicated both by the group's radio emission and morphological features, this mechanism could be present in NGC~5813 as well.

One of the conditions that should be met, as long as cooling via uplift is adiabatic, is that the entropy of the cool gas found at larger radii is similar to the entropy of the gas found in the centre of the group, which is the most likely point of origin. For the purposes of our analysis, we estimate the pseudo-entropy ($S$) for a double-phase gas, following \cite{Xue04}, as:
\begin{equation}\label{S}
    S=\text{exp}\left[\frac{fn_{e,c}\text{ln}S_c+(1-f)n_{e,h}\text{ln}S_h}{fn_{e,c}+(1-f)n_{e,h}}\right]
\end{equation}
where $S_h$ and $S_c$ are the pseudo entropies of the hot and cool temperature components respectively, defined as:
\begin{equation}
    S_j=\frac{kT_j}{n_{e,j}^{2/3}}.
\end{equation}
Here, $kT_j$ is the fitted projected temperature and $n_{e_j}$ the projected electron number density along the line of sight, for each thermal component for a given region, and $f$ the volume filling factor of the cool component. The volume filling factor is, in turn, defined as:
\begin{equation}\label{EM}
    \text{EM}_{c}=n^2_{e,c}f ~~~~ \text{or} ~~~~ \text{EM}_{h}=n^2_{e,h}(1-f)
\end{equation}
using the emission-measures of the cool and hot components respectively, with $n_{g,i}$ representing each component's gas number density. The emission-measures, approximated as:
\begin{equation}
    \text{EM}=\int n_e n_H dV \simeq n_e^2 V
\end{equation}
can be directly derived from each component's spectrum normalisation. We note that we approximate $n_e\sim n_H$, instead of $n_e=1.2 n_H$, as, due to the uncertainties in our volume estimates, this approximation can still provide us with a reasonable estimate of the order of magnitude. The volume derivation of each region, derived based on its geometry, is the following. For all the regions defined in \autoref{azimuthal_analysis} with $r>3~\rm{kpc}$, including the two surrounding atmospheres (bins 7 and 8, see \autoref{fig:central}) in the central region,  we assume that the gas between the two spheres defined by the maximum and minimum radii of the pixels in a given bin is the sole contributor to the detected emission \citep[e.g.][]{Henry04,Mahdavi05b}. As a result, we can derive the region's volume as
\begin{equation}
    V=\frac{2}{3}SL
\end{equation}
where $S$ is the projected region's area and $L=2\sqrt{R^2_{\text{max}}-R^2_{\text{min}}}$ the region's depth. For the inner cavities, we choose to approximate them using a spherical volume element instead, making a similar assumption regarding the longest line of sight, assuming it is equal to the sphere's diameter. For the four rims, we choose to not derive a volume estimate, due to the increased complexity and subsequent uncertainties in approximating the volume of a filament between two ellipsoids, and therefore they are excluded from our pseudo-entropy calculations.\\
Our final assumption, regarding the calculation of the pseudo-entropy, has to do with the gas thermodynamics, where we assume that the cool and hot gas phases are in local pressure equilibrium. This implies that
\begin{equation}\label{th_eq}
    n_{e,c}kT_c=n_{e,h}kT_h
\end{equation}
where $kT_c$ and $kT_h$ the temperature of the cool and hot gas respectively. Combining (\ref{EM}) and (\ref{th_eq}), one can solve for the volume filling factor, found to be:
\begin{equation}
    f=\frac{b}{1+b},~~~~b=\frac{\text{EM}_c}{\text{EM}_h}\left(\frac{kT_c}{kT_h}\right)^2
\end{equation}

\begin{figure}
    \centering 
    \includegraphics[width=\linewidth]{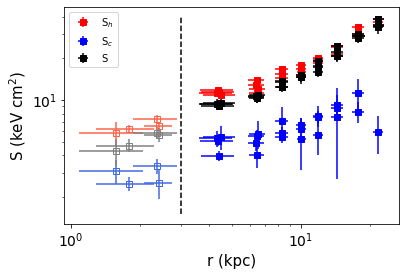}
\caption{Radial profiles of the double-phase gas pseudo-entropy ($S$), including the pseudo-entropy estimates of the hot ($S_h$) and cool ($S_c$) gas components. We present the values for all radial bins, including the two cavities and central surrounding atmosphere bins. The dashed line indicates a radial distance of $r=3~\rm{kpc}$.}
\label{fig:S_prof}
\end{figure}

In \autoref{fig:S_prof} we present the radial profiles of the double-phase gas pseudo entropy, as well as the individual pseudo-entropy profiles for the two thermal components. The radial trends indicate a clear overlap between the values of the pseudo-entropy of the hot gas in the center of the group and that of the cool gas at $r\sim 10~\rm{kpc}$, suggesting that adiabatic cooling via uplift is, in principle, possible.

However, if the observed cool gas were to uplifted, the amount of displaced gas and the energy required to do so should remain within physically plausible limits. To check this, we measured the total amount of cool gas found at a galactocentric radius $r>3~\rm{kpc}$ by summing over each radial bin, following \citet{Simionescu08}:
\begin{equation}
    M_c\approx \braket{m_i}\sqrt{\text{EM}_cV_c/f_{ei}}\approx m_p\sqrt{\text{EM}_cV_c}
\end{equation}
where $\braket{m_i}$ is the average ion mass per electron, $m_p$ the proton mass, $f_{ei}$ the electron to ion number ratio of $\sim1$, following a similar reasoning as before, and finally $\text{EM}_c$ and $V_c$ are the emission measure and volume for the cool gas, accounting for the bin's cool gas volume filling factor. A similar methodology has been applied for the estimation of the total (cool+hot) gas mass found in the group's central 3 kpc. As to avoid large uncertainties in the associated volume estimates, a larger binning has been implemented by regrouping the central cavities and rims (bins 1-6 in \autoref{fig:central}) into a single ellipsoid. While such a scheme might bias our spectral fit, given that said regions have different physical properties, it allows us to more accurately measure the total gas mass in the centre.

Based on our analysis, we derived a total mass for the cool (presumably uplifted) gas of $M_c\approx 2 \times 10^8 ~M_\odot$, or $\sim 2$ times the gas mass found in the group's centre. This amount is comparable to, although generally smaller than, the amount of uplifted gas found in M87 \citep[][]{Werner10,Simionescu18} and Hydra A \citep[][]{Simionescu09} with $6-9 \times 10^8~M_\odot$ and $1.6-6.1\times10^9~M_\odot$ respectively. However, the relative value appears surprisingly high; if the cool gas at larger radii were to originate solely from uplift, this would imply that twice more gas is transported outwards than what is left in the group's central region. We also estimated the gravitational potential energy of the cool gas, which could provide us with a rough estimate of the energy required to displace it. Using the group's pressure profile in order to derive $g$, we measure the total gravitational potential energy of the cool gas ($U=M_cgr) $ based on the amount of cool gas found at a given radius. We find a potential energy of $U\simeq 2.26\cdot 10^{57}~\rm{ergs}$, which is comparable to the estimated energy released by the the outburst that generated the intermediate cavities \citep[][]{Randall11}. This indicates that said outburst could uplift the observed cool gas from the centre of the group along the direction of its lobes. The larger ratio of uplifted to central gas mass, compared to clusters of galaxies, may reflect the ability to move larger amounts of gas due to the group's shallower gravitational potential well.

Overall, cooling via uplift appears to be the most likely origin of the cool gas in NGC~5813, as it is able to reliably explain the extent and directionality of its distribution. However, in-situ cooling can not be explicitly excluded, as it can still be present in the NGC~5813 group's centre. Additional thermal processes, such as conduction, could also be present, as a strictly adiabatic uplift would imply a decrease in the cool gas temperature as a function of radius, where, in our case, a small gradient in the temperature rising with radius is found instead. This possibility can be further substantiated by the fact that an extended H$\alpha$ emission has been found to be co-spatial with the location of the inner and intermediate cavities \citep[][]{Randall11}. This suggests that, alongside the X-ray emitting gas, some much cooler atomic, and possibly molecular, gas has been uplifted from the centre as well, that could act as an effective heat sink that cools down some of the gas in place. Such a scenario could explain the surprisingly high mass of uplifted gas observed, as some of the cold gas would not be originating from the centre.

It is also important to note that a double temperature model might not be sufficient in accurately describing the multi-temperature structure of the IGrM and, therefore, its exact origin. A third temperature component, or even a continuous distribution of temperatures might be necessary in order to accurately describe it. Such an examination cannot be performed due to the current instrumental limitations, but should be further investigated when high resolution spectra become available.

\subsection{Metal Mixing}
\begin{figure}
    \centering 
    \includegraphics[width=0.9\linewidth, height=6cm]{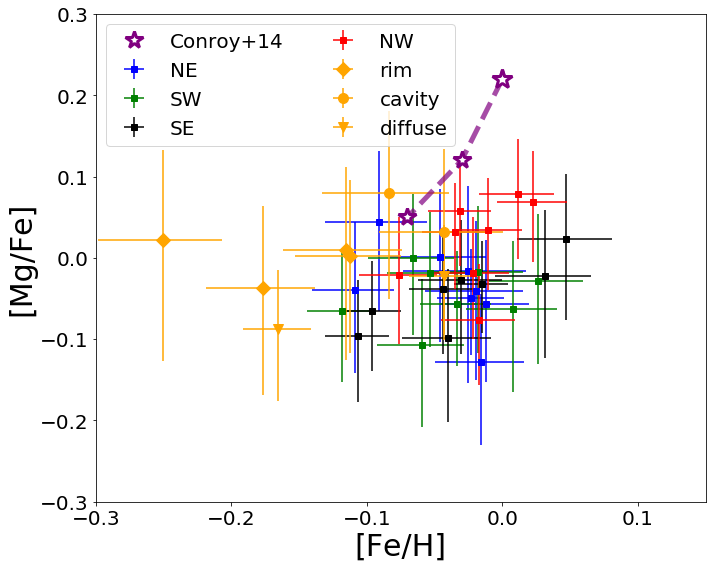}
    \medskip
    \includegraphics[width=0.9\linewidth, height=6cm]{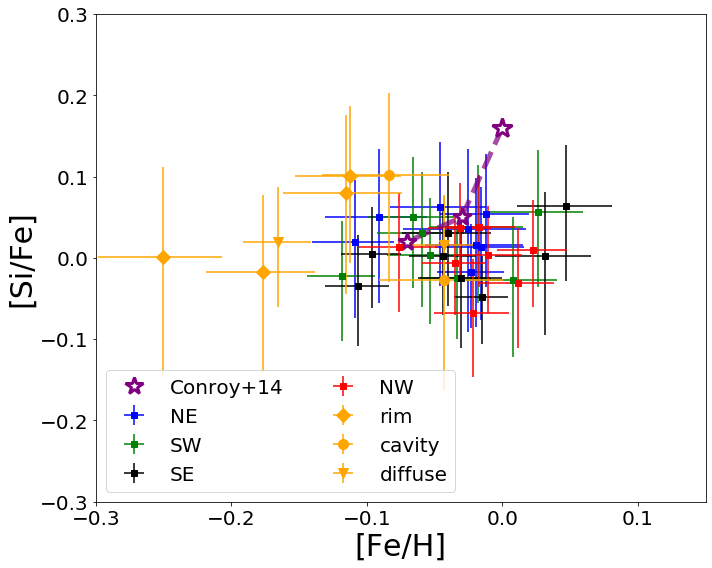}
    \medskip
    \includegraphics[width=0.9\linewidth, height=6cm]{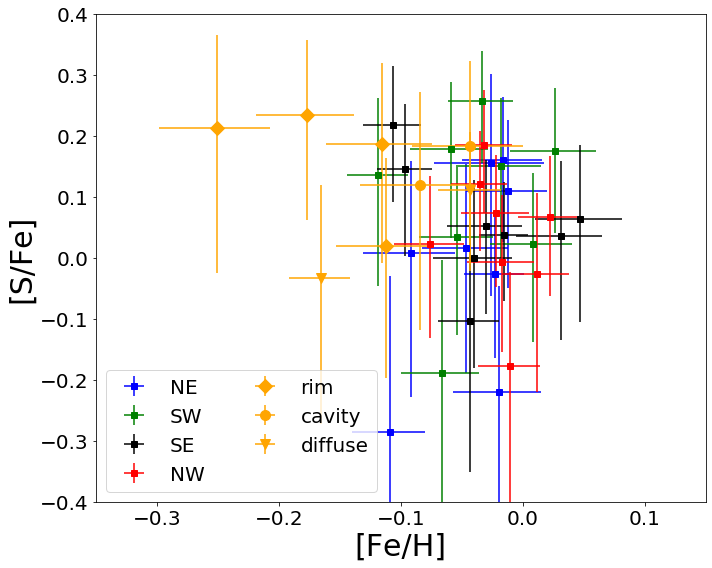}
\caption{Elemental abundance over iron ratios as a function of iron abundance, for all radial bins, including all the central bins. For Mg and Si, where data were available, we include the stellar abundance estimates for massive elliptical galaxies, according to \citet{Conroy14}, for high, low and average velocity dispersion. The high velocity dispersion estimate is indicated more prominently, as it is the likeliest case for NGC~5813.}
\label{fig:element_rat}
\end{figure}

Another interesting aspect of NGC~5813's IGrM is the overall uniform spread of elements observed. As previously indicated, the abundances of all elements examined, namely Fe, Mg, Si and S, remain relatively close to Solar, independent of both radius and azimuth. However, as indicated by \autoref{fig:element_rat}, the derived abundances suggest a much lower $\alpha$-element over Fe ratio compared to the stellar abundances of typical BCGs, as indicated by optical spectroscopy. For example, a massive galaxy, such as NGC~5813, is expected to have a Mg/Fe ratio of $\sim 1.66$ and Si/Fe ratio of $\sim 1.45$, corresponding to the highest $\sigma_v$ early-type galaxies probed by \citet{Conroy14}, both of which remain significantly higher than our derived ratios.

The elemental abundances also appear to be independent of the thermodynamics of the gas. As we have previously shown, there is no apparent correlation between them and the presence of cool gas in the IGrM. In fact, we find no correlation between the Fe abundance and the cool gas content, indicated via the volume filling factor, with a Pearson coefficient $\rho=0.279$. Similarly, the Fe abundance is independent of the pseudo-entropy of the gas, with little-to-no anticorrelation between the two variables, with a Pearson $\rho=-0.338$, unlike previous studies of clusters, where a correlation between the two has previously been reported \citep[e.g.][]{Ghizzardi14}.

Overall, our results indicate a well-mixed IGrM, with a uniform elemental abundance across all directions, and lacking a central enhancement peak, compared to many other relaxed systems \citep[e.g.][]{dePlaa17,Simionescu09, Sun12, Mernier17, Lovisari19, Gastaldello21}. This indicates that the IGrM of NGC~5813 must have been very turbulent in the past, facilitating an efficient mixing. The asymmetry in the Fe abundance for the two central surrounding atmosphere bins (bin 7 and 8), might also indicate a scale dependence on said mechanism, with the central $r<3~\rm{kpc}$ bins having not reached uniformity yet.

 One mechanism that could be responsible for such mixing is the AGN feedback itself, with its effects being magnified due to the lower mass of the system. For example, cavities, while buoyantly moving upstream of the uplifted gas, can generate turbulence capable of redistributing any elements present in said gas \citep[][]{Churazov01}. However, while in clusters the effect remains localised, in less massive systems the generated turbulence could potentially grow larger and therefore diffuse any abundance enhancement due to uplift. Similarly, gas uplift becomes a far more efficient process, due to the shallower potential wells of the less massive systems, moving a much larger fraction of the enriched central gas outwards. This means that the central abundance peak is significantly reduced, resulting in the hampering of the central enhancement peak, similar to what we see in NGC~5813. Evidence of gas fallback, referring to uplifted gas falling back to the group's centre, can also be found in the cool gas distribution, with the prominent SE feature and the less prominent SW one serving as potential examples of uplifted gas turning back towards the central galaxy in a fountain-like fashion. These large amounts of cool gas falling back to the centre can also create large-scale wakes with their movement through the IGrM, further contributing to the mixing of elements.
 
Another possible mechanism capable of generating such large scale turbulence in the IGrM is sloshing. If gas is displaced from the centre of the gravitational potential well and, in the process, gains angular momentum, it will not fall back but rather continue to move around the system. This phenomenon is referred to as sloshing and its presence can be identified by a series of cool fronts or substructures such as spiral brightness residuals or "arcs" of enhanced emission on opposite sides of the group, associated with the gas' movement\citep[][]{Markevitch07}. As a result, due to its scale, sloshing can be a very efficient way to distribute metals across the entire IGrM \citep[][]{Simionescu10}. In relaxed clusters \citep[e.g.][]{PaternoMahler13,Ascasibar06,Sanders16b} sloshing is thought to be driven by minor interactions with subclusters outside of them.

The primary reason to speculate that sloshing may take place in NGC~5813 has to do with the morphology of the cool gas. As we have pointed out, both the prominent SE feature and the less prominent SW one appear to be rotating in the same (counter-clockwise) direction. However, the system's overall morphology is fairly regular, with the three generations of cavities remaining remarkably aligned, and no signs of asymmetrically placed surface brightness edges. The lack of said edges is interesting, since, as indicated by \citep[][]{Roediger11}, some sharp morphological features should be visible, regardless of the projection angle. Therefore, if a sloshing event is responsible for the observed well-mixed abundance distribution, it must have occurred a long time ago in the group's history, so that any sharp features in the X-ray brightness have been diffused and broken down by instabilities by now. While this mechanism could produce large scale turbulence, this mechanism is less likely than AGN feedback, as it would constitute a manifestation of sloshing that is otherwise not commonly observed.

Overall, both the uniformity of the elemental abundance observed in NGC~5813 and the morphology of the cool gas indicate the presence of complex dynamical processes in the galaxy group, despite what its regular morphology initially suggests.

\section{Conclusions}\label{sec:conclusions}
We have re-examined the properties of the IGrM for the NGC~5813 galaxy group, using archival deep Chandra observations. The main difference in the analysis presented here, compared to previously published work using the same data set, is that we have implemented a double temperature model in our spectral fitting. Because interactions with the AGN are known to lead to multiple temperature components in the hot atmospheres surrounding other BCGs, this method is critical for understanding the thermodynamical and chemical structure of the gas around NGC~5813, and mitigating significant problems related to the iron bias.
We now summarise our main results that remain consistent even when different atomic libraries or model parameterisations are used, as indicated by our analysis of the systematic uncertainties (see \autoref{systematics}).
\begin{itemize}
    \item[-]Our results indicate the presence of an extended cool gas component with a morphology remarkably associated with the direction of the cavities/jets of NGC~5813. The bulk of the cool gas traces the intermediate pair of AGN-inflated cavities, while a small fraction of it seems to be bending away from them in a counterclockwise fashion, and possibly falling back towards the group center.
    \item[-]The bulk of the cool gas is found in regions where the cooling time to free fall time ratio of the hot atmosphere is much greater than 10, leading us to conclude that in situ thermal instabilities are unlikely to be the main mechanism responsible for this structure. Instead, the cool gas is most likely uplifted from the group's centre, during "outburst" events in the central AGN's history. This is corroborated by the entropy of the cool gas, that is very similar to the entropy within the central region (<3~kpc) of the group. The total mass of cool gas observed ($2\times10^8~M_\odot$) is comparable to other known examples of gas uplift, but amounts to a surprisingly high fraction of the total gas mass within the central 3~kpc. However, the potential energy of the cool gas ($2.26\times10^{57} \rm{ergs}$) remains comparable to the mechanical energy needed to inflate the intermediate cavities. This suggests that gas uplift by the AGN may be remarkably efficient in NGC~5813, although we cannot exclude the possibility that local thermal instabilities triggered by the AGN interaction in situ may have also contributed to further increasing the cool gas mass.
    \item[-] Our abundance estimates suggest an overall uniform distribution across the galaxy group, lacking a central enhancement peak, commonly found in relaxed systems. The elemental distribution appears to be on average Solar, independent of galactocentric radius, azimuth, cool gas content and pseudo-entropy. 
    \item[-] Our results suggest that complex dynamical processes are present in the galaxy group, as indicated by the efficient metal mixing found, despite the group's regular morphology. We speculate that turbulence driven by the amplified effects of AGN feedback, due to the lower mass of the system have produced the observed abundance uniformity. However, though less likely, a past sloshing event may have aided this mixing, and could also be responsible for the morphology of the cool gas plumes that seem to bend in a counterclockwise fashion towards larger radii.
\end{itemize}
While the overall trends, or lack thereof, found in this analysis highlight the large scale properties of NGC~5813's IGrM, our ability to distinguish between possible scenarios regarding the origin of the cool gas and the dynamical processes present is primarily limited by our resolution. Future missions, such as XRISM and Athena, due to their higher spectral resolutions would be able to better resolve both the cool gas distribution and dynamics, such as turbulence, allowing for a more detailed study of the diffuse gas in galaxy groups.

\section*{Acknowledgements}
We thank Simona Giacintucci for sharing the GMRT image of NGC 5813. We thank Scott Randall for discussions that helped to improve this manuscript. We also thank the anonymous referee for their helpful comments. The scientific results reported in this article are based in part on data obtained from the \textit{Chandra} Data Archive. A.S. acknowledges funding from the Women In Science Excel (WISE) programme of the Netherlands Organisation for Scientific Research (NWO), and thanks the Kavli IPMU for their continued hospitality. SRON Netherlands Institute for Space Research is supported financially by NWO. 

 \section*{Data Availability}
The data underlying this article were accessed from Chandra Interactive Analysis of Observations' (CIAO) archive. The GMRT data used in this article were provided by Simona Giacintucci by permission. Data will be shared on request to the corresponding author with their permission. The derived data generated in this research will be shared on reasonable request to the corresponding author.


\bibliographystyle{mnras}
\bibliography{ms}



\appendix
\section{Thermodynamic properties}\label{thermodynamics}
\begin{figure*}
    \centering 
    \includegraphics[width=\linewidth]{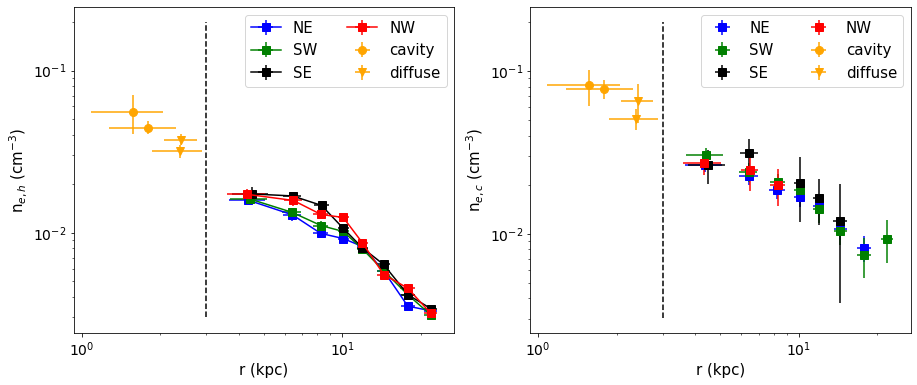}
\caption{Radial profiles of the hot (left) and cool (right) gas component's electron density for all 4 directions, including the two inner cavities and central surrounding atmosphere bins. The dashed line indicates a radial distance of $r=3~\rm{kpc}$.}
\label{fig:ne_prof}
\end{figure*}

\begin{figure}
    \centering 
    \includegraphics[width=\linewidth]{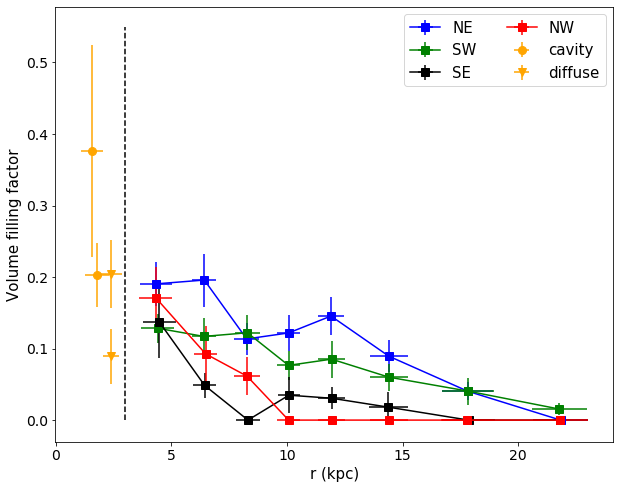}
\caption{Radial profiles of the cool gas volume filling factor (f) for all 4 directions, including the two inner cavities and central surrounding atmosphere bins. The dashed line indicates a radial distance of $r=3~\rm{kpc}$.}
\label{fig:f_prof}
\end{figure}

In this section we present the radial profiles of the intermediate thermodynamic quantities used in the calculations of the pseudo-entropy profiles, namely of the cool gas volume filling factor ($f$) and the electron densities of the hot ($n_{e,h}$) and cool ($n_{e,c}$) gas components, presented in \autoref{fig:f_prof} and \autoref{fig:ne_prof} respectively.

Regarding the cool gas volume filling factor, we observe an overall radial decrease across all four directions, representing a radial decrease in the cool gas content. However, the two lobe directions (i.e. NE and SW) retain a significant cool gas component ($\sim 10\%$ filling factor) for longer distances than the two off-lobe directions, consistent with our derived distributions. More notably, we observe two instances where the cool gas component increases. The first occurs in the SE direction, at a distance of $\sim 10~\rm{kpc}$ and is most likely associated with the extended cool gas feature, that appears to be turning back towards the centre, seen in \autoref{fig:temp}. The second one occurs in the NW direction at $\sim 4~\rm{kpc}$, and is evident when comparing NW's first radial bin to the NW central surrounding atmosphere (bin 7). Since the latter overlaps almost entirely with our azimuthal cuts of the NW direction, we can consider it as the 0th bin in that direction's radial profile. As a result, we have an almost $100\%$ increase in the cool gas component, from $f=0.1$ to $\sim 0.2$, indicating a cool gas injection (or pile-up) at the given scale.

\section{Systematics}\label{systematics}

In this section we discuss the robustness of our results, by examining the extent to which they depend on the model parameterisation and the atomic library used in our analysis. Regarding the parameterisation of the model, we first examine the effects of the absorption component on our results. In our main analysis, we have assumed a fixed absorption along the line of sight, with a column density of $N_H=4.37\times10^{20} \rm{cm}^2$. However, X-ray absorption itself can exhibit a spatial gradient, and there are known cases where the X-ray-fitted column density does not match the value proposed by the LAB measurements \citep[e.g.][]{Mernier16a,dePlaa17}. In order to accommodate for such variations, we allow the absorption parameter to assume values within the $[2.185~,~8.74] \times 10^{20} \rm{cm}^{-2}$ range, representing half and double the original value used for the column density originally. The allowed range of values is conservatively high, illustrating the maximum uncertainties that could be introduced due to absorption in our abundance estimates.

\begin{figure}
    \centering 
    \includegraphics[width=\linewidth]{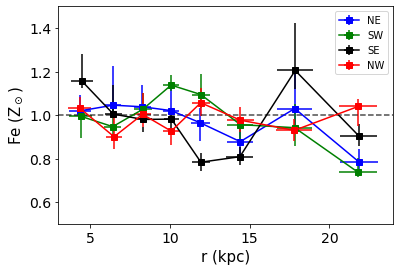}
\caption{Radial profiles of the Fe abundance for all 4 directions using a double temperature model with a free absorption component. The dashed line indicates Solar abundance.}
\label{fig:var_nh}
\end{figure}

In \autoref{fig:var_nh} we present the abundance radial profiles for all 4 directions using a double temperature model with a free absorption component. It is evident, that the radial trends remain consistent with both our previous results (see \autoref{fig:abund_radial}) and a uniform distribution of elements across all azimuths, although the abundances in the free nH case are slightly higher than those with a fixed nH. Overall, the persistence of a similarly well mixed IGrM across all azimuths, further validates our initial results.

We also examine the effect of including a second thermal component on the chemical structure of the IGrM. As previously stated, Fe-L complex spectroscopy could be subjected to effects, such as the Fe-bias \citep[][]{Buote99}, that can impact our understanding of the actual chemical structure of groups. Since we have argued that such effects could explain the inconclusive literature regarding the effects of the AGN feedback on the chemical enrichment of less massive systems, we have reanalysed our abundance profiles ignoring the multi-temperature structure of the gas.

\begin{figure*}
    \centering 
    \includegraphics[width=0.4\linewidth]{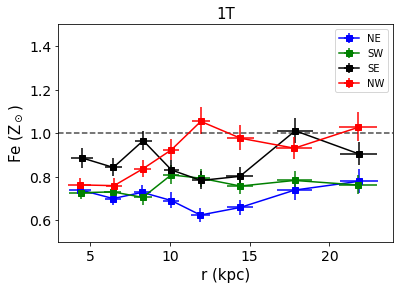}
    \includegraphics[width=0.4\linewidth]{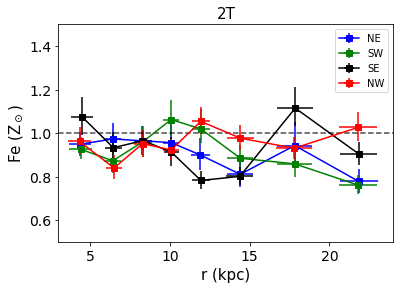}
\caption{Fe abundance radial profiles for all 4 directions, using a single (left) and a double temperature model (right). The dashed line indicates Solar abundance.}
\label{fig:1T}
\end{figure*}

In \autoref{fig:1T} we present the Fe abundance across all 4 directions, based on a single temperature model, compared to the previously derived double temperature ones. It is evident that the Fe-bias significantly impacts the two on-lobe directions (i.e. NE and SW), where the majority of the cool gas has been found, with their abundances being systematically underestimated. However, the off-lobe directions (i.e. SE and NW) also appear to be impacted at distances where a prominent cool gas contribution, such as the SE feature and the central radial annuli, is found. By ignoring the multi-temperature nature of the gas, the distribution of elements appears to be asymmetric, with an increased Fe abundance across the off-lobe directions, with a significantly metal poor centre, contradicting any notion of a well-mixed IGrM. While, as we have pointed out, a strictly double temperature gas might not fully describe the actual multi-temperature structure, with a continuous distribution of temperatures being more likely, completely ignoring said multi-phase structure severely impacts our understanding of the impact AGN feedback has on the chemical evolution of low mass systems.

The final source of systematic uncertainties, regarding the parameterisation of our models, that we choose to examine is related to the treatment of the abundances of individual elements. In our analysis, we have assumed that all elements, with the exception of Fe, Mg, Si and S, had solar abundances that were not allowed to vary (Model 0). However, elements such as Ni, Ne and O, while not as prominent as Fe, still have emission lines in the $0.6-3 ~\rm{keV}$ spectral window and it is therefore worth examining their impact when incorporated in our models. For this reason, we choose to implement a variation of our previously used model, in which the contribution of those three elements is incorporated in the estimate of the Fe abundance, by coupling them to it (Model 1). The remaining elemental abundances are still set to a fixed 'Solar' abundance. For the final model variation, we choose to expand on the contribution that the rest of the elements have on the abundance estimates by coupling all elements heavier than B, excluding Mg, Si and S, to the Fe abundance estimate (Model 2). One can think of this model parameterisation as a proxy to the APEC model, where a single abundance estimate is derived while elements below C are assumed to have Solar abundances, with the exception that the secondary contributors (i.e. Mg, Si and S) are accounted for independently.

\begin{figure*}
    \centering 
    \includegraphics[width=0.8\linewidth]{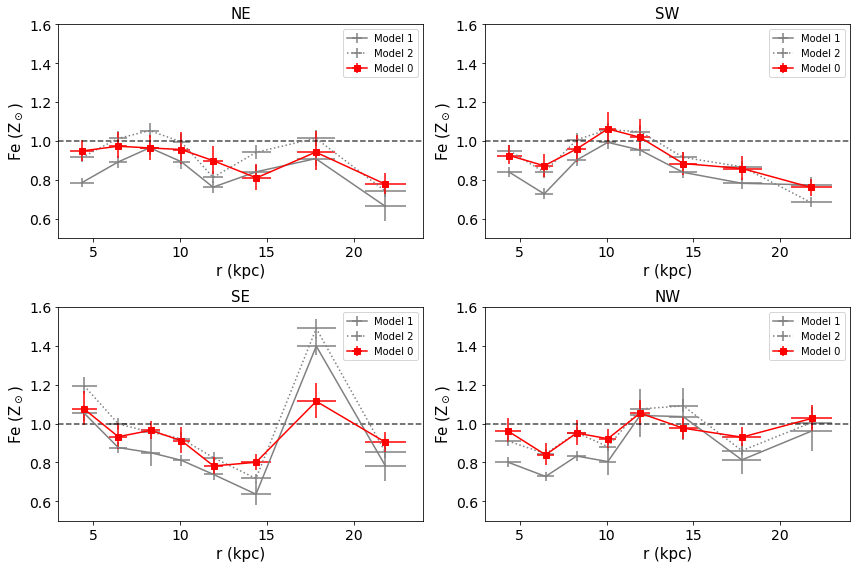}
\caption{Radial Fe abundance profiles, using different model parameterisations, for each direction. The description of each model can be found in the text. The dashed line indicates Solar abundance.}
\label{fig:vapec_models}
\end{figure*}

In \autoref{fig:vapec_models} we present the Fe abundance radial trends, as this is the primary abundance tracer, for all 4 directions, using our previously described model variations. Our results indicate only small deviations from our initial trends, with the majority of estimates remaining within the 1$\sigma$ uncertainties of the fiducial model presented in the main body of the paper.

\begin{figure*}
    \centering 
    \includegraphics[width=0.8\linewidth]{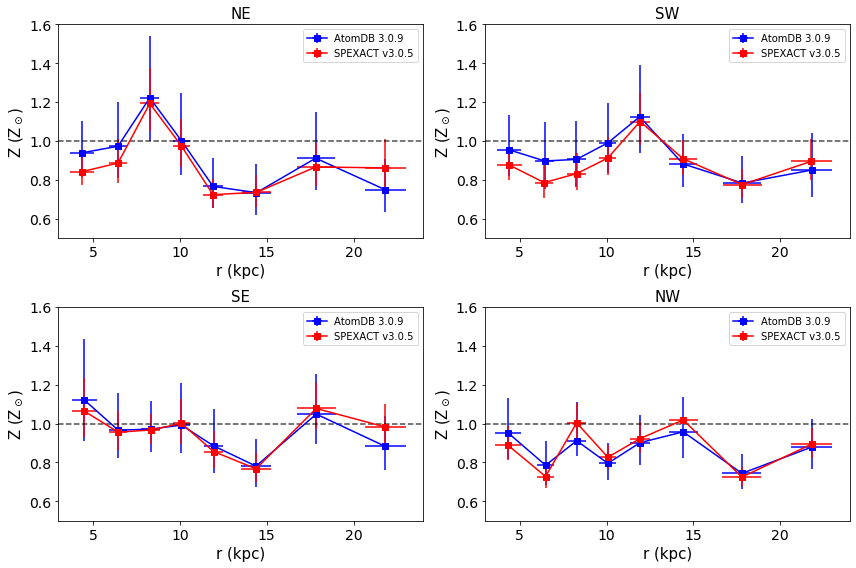}
\caption{Radial abundance profiles, using different atomic libraries, for each direction. The dashed line indicates Solar abundance. Error bars indicate 90\% confidence intervals.}
\label{fig:spex_abund}
\end{figure*}

\begin{figure*}
    \centering 
    \includegraphics[width=0.8\linewidth]{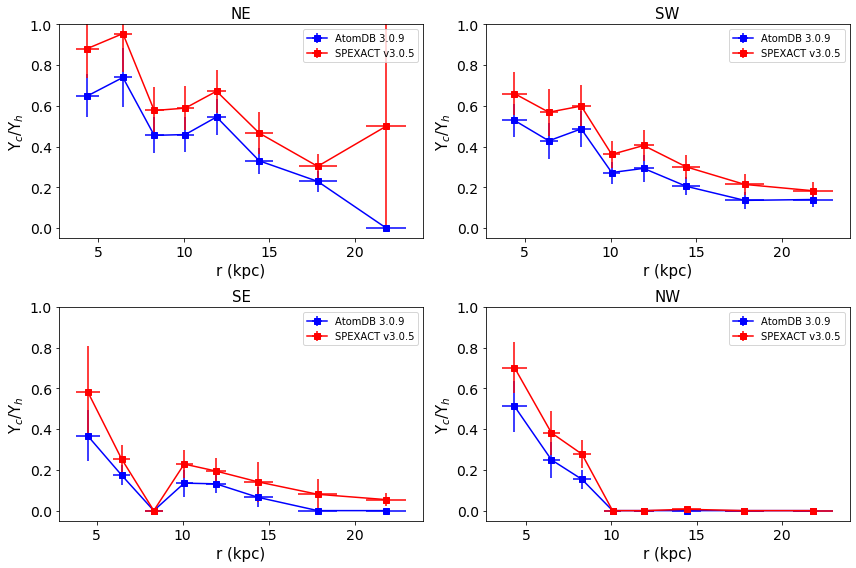}
\caption{Radial component normalisation ratio profiles, using different atomic libraries, for each direction.}
\label{fig:spex_norm}
\end{figure*}

Another major source of systematic uncertainties is related to the atomic library itself. Previous studies have indicated that discrepancies in the elemental abundances can arise when different atomic libraries are used for the spectral fitting \citep[e.g.][]{Mernier18c,MernierdePlaa18,Mernier18b,Mernier20}. For this reason we choose to examine if our results, derived using the AtomDB 3.0.9 database implemented in XSPEC, remain consistent if the SPEXACT v3.0.6 (SPEX Atomic Code and Tables) library is used instead. In \autoref{fig:spex_abund} and \autoref{fig:spex_norm} we present the abundance and component normalisation ratios for all 4 directions for a double temperature model, between AtomDB 3.0.9 and SPEXACT v3.0.6. Based on our analysis, we find little to no difference on the derived abundances, indicating the robustness of our results. However, we do see a systematic offset in the relative contribution of the two thermal components with SPEXACT indicating a more prominent second temperature component, while within error bars. The main reason behind this offset appears to be the differences in the implementation of the emission lines in the two atomic codes, where the under-prediction of the emissivities of certain lines within the $0.6-3 ~\rm{keV}$ energy band could slightly boost the strength of the secondary thermal component without significantly affecting the model's derived overall abundance.

\bsp	
\label{lastpage}
\end{document}